# Transition Metal-Anchored Newly Synthesized Aza-Triphenylene Monolayer for Efficient Hydrogen Evolution Reaction: Insights from DFT

Soumendra Kumar Das[1,2], Jagannath Pradhan[2], Prasanjit Samal[3], Sridhar Sahu[2] and Brahmananda Chakraborty[4,5]

[1]*Department of Chemistry and Chemical Engineering, Inha University, Incheon, 22212, Republic of Korea*

[2]*Department of Physics, Indian Institute of Technology (Indian Institute of Mines) Dhanbad, Jharkhand-826004, India.*

[3]*School of Physical Sciences, National Institute of Science Education and Research (NISER) Bhubaneswar, An OCC of Homi Bhabha National Institute, Bhubaneswar, 752050, Odisha, India.*

[4]*High Pressure Crystallography Section, Bhabha Atomic Research Centre (BARC), Trombay, Mumbai 400085, India*

[5]*Homi Bhabha National Institute, Anushaktinagar, Mumbai 400094, India*

Corresponding author: *brahma@barc.gov.in*

**Abstract**

First-principles calculations were employed to explore the hydrogen evolution reaction (HER) activity in two-dimensional aza-triphenylene monolayers functionalized with transition metals. Among the various metals studied, Co, Ru, and Y demonstrate outstanding catalytic efficacy, lowering the hydrogen adsorption free energy from -0.78 eV for pure aza-triphenylene to 0.05, 0.20, and 0.06 eV, respectively. Volcano plot analysis corroborates their enhanced HER activity. Ab initio molecular dynamics simulations confirm thermal stability, while the electronic structure and charge transfer analysis reveal the strong catalyst-surface interaction. Metal decoration also improves the carrier mobility, particularly for Ru-functionalized aza-triphenylene. The climbing-image nudged elastic band calculations for Co-functionalized aza-triphenylene exhibit activation barriers of 1.61, 2.29, and 5.88 eV for the Volmer, Heyrovsky, and Tafel reactions, respectively, highlighting a preferred Volmer–Heyrovsky route with the Heyrovsky being the rate-determining step. These findings highlight the potential of transition-metal-functionalized aza-triphenylene as a promising platform for developing efficient HER electrocatalysts.

## Introduction

The pursuit of clean, sustainable, and renewable energy sources has become a global priority in order to address the pressing challenges of climate change and environmental degradation. Hydrogen is widely considered as an optimal energy carrier due to its elevated gravimetric energy density and its environmentally benign combustion byproduct -water[1]**.** Among the numerous strategies for hydrogen production, electrochemical water splitting emerges as a promising technology, with the hydrogen evolution reaction (HER) playing a pivotal role in its implementation. Electrochemical water splitting, which includes the oxygen evolution reaction (OER) at the anode and the hydrogen evolution reaction (HER) at the cathode, provides a direct and environmentally friendly route for hydrogen production. Although platinum remains the benchmark catalyst for HER because of its nearly thermoneutral hydrogen adsorption free energy ($\Delta G_{HER}$), its limited availability and exorbitant price restrict its widespread applications. Consequently, designing earth-abundant, cost effective, and efficient HER catalysts has become a central research focus[1-3]**.**

Two-dimensional (2D) materials have arisen as prominent options for HER electrocatalysis because of their large surface area, customizable electrical characteristics, and ample active sites. Their reduced dimensionality ensures efficient charge transport, while structural flexibility allows for band-edge engineering to optimize catalytic performance[2]. Recent advances include single-atom-decorated graphyne, which demonstrates multifunctional electrocatalytic activity toward HER, OER, and ORR, showcasing the potential of atomically dispersed active sites on carbon-based 2D frameworks. Similarly, heterostructures such as g-$C_3N_5$/graphene have been shown to promote HER via enhanced interfacial charge transfer and improved separation of photogenerated carriers, thereby eliminating the need for expensive noble metals[2, 3]**.**

Another important class of emerging 2D HER catalysts is transition-metal-based chalcogenides and tellurides. For example, $Fe_2Te_2$ monolayers doped with transition metals have been reported to activate otherwise inert basal planes, significantly lowering hydrogen adsorption barriers and enabling high HER activity[4]**.** These discoveries emphasize the importance of dopant engineering in tuning the intrinsic activity of layered materials. Light-element-based 2D systems, especially those incorporating boron, are increasingly recognized for their electronic and catalytic tunability. Boron monoxide (BO), long known but only recently structurally resolved, exemplifies this potential. BO consists of layered $B_4O_2$ ring units arranged in disordered stacking, resolving a decades-old structural question[5]. Parallel computational studies revealed that BO monolayers are indirect-gap semiconductors (~3.8 eV, HSE06), mechanically robust, and thermally stable at elevated temperatures. Decoration of BO with alkali and alkaline-earth metals (Li, Na, K, Ca) significantly enhances hydrogen affinity, achieving gravimetric storage capacities up to 11.75 wt% [6, 7].

Beyond inorganic 2D lattices, covalent organic frameworks (COFs) provide a modular approach to designing HER catalysts. Their crystalline and porous architectures offer extended π-conjugation and customizable active sites. A notable example is the aza-triphenylene-based COF decorated with scandium, which demonstrated a storage capacity of hydrogen around 8.43 wt% and favourable hydrogen binding energies, suggesting potential for HER electrocatalysis[8]. Further extending this concept, lithium decoration on aza-triphenylene COFs enabled hydrogen uptake up to 9.49 wt% with optimal adsorption energies (~0.30 eV/$H_2$), balancing storage and release under ambient conditions[9]. These studies illustrate the versatility of COFs as platforms that can be tailored for either hydrogen storage or for HER catalysis. Carbon-based allotropes such as biphenylene nanoribbons (BPRs) exhibit excellent HER activity, with favourable hydrogen adsorption free energies, high carrier mobility, and kinetically feasible pathways for adsorption/desorption, indicating their potential as efficient HER catalysts[10]. Boron phosphide (BP) has gained attention for its direct 1.35 eV bandgap, ultra-large electron mobility (~$8.46 \times 10^4$ $cm^2$ $V^{-1}$ $s^{-1}$), and favourable band edge positions that enhance charge transfer and HER activity, with tensile strain providing robust catalytic activity across a wider pH range[1].

Nickel-based materials like $Ni_2B_5$ and $NiC_3$ exhibit ultralow hydrogen adsorption barriers (~0.096 and 0.018 V), comparable to platinum, due to dual active centres formed by nickel's interaction with boron and carbon, enhancing catalytic efficiency. Further analysis of nickel monolayers confirmed superior overpotentials and optimal adsorption sites. Antimony-doped nickel catalysts stabilize intermediates and lower energy barriers, improving both HER and OER performance, demonstrating how elemental substitution enhances reaction pathways. Similarly, nitrogen-doped graphene frameworks show that solvation and dopant concentration stabilize adsorption sites, boosting oxygen reduction kinetics[11-13].

A comprehensive understanding of carrier mobility and charge transport is critical for catalyst optimization. A review showed that doping, structural defects, and mechanical strain influence carrier mobility in 2D materials. Their insights provided guidelines for designing electrocatalysts with enhanced conductivity and efficient electron transfer pathways, directly impacting hydrogen evolution rates[14]. Solvation effects and interfacial interactions have also been systematically investigated. A study analysed how explicit water solvation and nitrogen site distribution influence the adsorption and reaction pathways in nitrogen-doped graphene, emphasizing the need for controlled environments to achieve optimal catalytic activity[13]. Similarly, another study explored transition-metal-anchored $C_2N$ monolayers and demonstrated that metal-support interactions significantly affect hydrogen adsorption energies and catalytic activity profiles, paving the way for bifunctional electrocatalysis in hydrogen and oxygen evolution reactions[15].

Hydrogen storage systems such as $BC_3N_2$ frameworks have also been computationally evaluated. First-principles simulations revealed that these frameworks not only support substantial hydrogen uptake but also exhibit reversible adsorption properties with low binding energies, ensuring effective storage and catalysis integration[16]. These results complement efforts to design catalysts that meet the dual demands of hydrogen generation and storage. Benchmarking oxygen reduction mechanisms on platinum and alloys provides insights into evaluating emerging catalysts, while adsorption energetics and structural stability are key for long-term performance. Transition metal carbides and nitrides mimic noble metals with improved stability and lower cost, and defect engineering offers strategies to optimise catalytic surfaces for activity and durability[17-20].

Aza-triphenylene is a promising two-dimensional nitrogen-rich carbon-based material due to its delocalized π-conjugated framework, intrinsic electronic properties, and chemically active nitrogen sites, which make it attractive for energy-related applications[9]. However, the pristine material may not provide sufficiently strong and tunable interactions with adsorbates[21], motivating the incorporation of suitable metal atoms to modify its electronic structure and surface reactivity. It has been found that the HER and OER activities of 2D materials exhibit a significant enhancement through hetero atom doping[22] . In particular, the design and development of single-atom catalysts for enhancing the HER and OER activities have been a serious focus for many years, which can overcome the shortcomings of traditional catalysts in terms of optimal atom use and manipulation of electronic structure[23]. Wu et al. analysed the effect of transition metal adsorption on g-$C_3N_4$ and indicated that such systems are effective bifunctional electrocatalysts, as the introduction of transition metals can markedly diminish the overpotentials associated with the oxygen evolution and reduction reactions[24]. The adsorption of transition metals as a single-atom catalyst on the surface of $C_2N$ monolayer reduced the overpotential and enhanced the HER and OER catalytic performance [15]. Experimental investigations have revealed the enhanced electrochemical performance of shear-exfoliated Sb in relation to the HER, OER, and ORR when contrasted with other group-VA elements[25]. Therefore, it is very stimulating and interesting to examine the influence of single transition metal adsorption on the aza-triphenylene monolayer surface for efficient HER catalytic activities aiming at sustainable hydrogen production.

In the present work, we therefore systematically investigate pristine and transition-metal-decorated aza-triphenylene using first-principles calculations to elucidate their thermal stability, electronic properties, metal–framework interactions, and strain-dependent behaviour. Attention is given to the interaction of the decorated systems with hydrogen and their potential toward the hydrogen evolution reaction (HER), supported by adsorption energetics, electronic-structure analysis, charge-transfer characteristics, ICOHP calculations and carrier mobility to establish the underlying bonding mechanism. The thermal stability of the pristine and decorated aza-triphenylene monolayer was verified by ab-initio molecular dynamics simulations. The mechanism of HER for the most promising candidate was studied by the Climbing Image Nudged Elastic Band (CINEB) method under the Volmer, Heyrovsky and Tafel reaction schemes in an acidic medium. The charge transfer analysis was executed using the charge density difference and Bader charge calculations[26].

## Computational Methods

In the current study, we have used first-principles density functional theory calculations using a plane-wave basis set as implemented in the Vienna ab-initio Simulation Package (VASP)[27, 28]. The projector augmented method was used to analyse the interaction between the ions and electrons[29]. The Perdew, Bruke and Ernzerhof (PBE) parametrization based generalized gradient approximation (GGA) method was adopted to describe the exchange-correlation functional[30]. The kinetic energy cutoff for the plane wave basis set was set at 520 eV. The Brillouin zone sampling was performed using a $5 \times 5 \times 1$ Monkhorst–Pack K-mesh grid for the geometrical optimization. However, a comparatively dense k-mesh $7 \times 7 \times 1$, and $14 \times 14 \times 1$ were used for the ground-state energy and projected density of states calculations. For the transition metal-decorated AZA-COF, spin-polarised electronic structure calculations were considered. The Grimme-D3 van der Waals correction technique with the Becke-Johnson damping function was incorporated to precisely characterise the long-range van der Waals interactions. A 20Å vacuum layer was selected along the c axis to avoid interactions with the periodically repeated images. Due to the inadequacy of the PBE-GGA functional, we have used HSE-06 hybrid functional to accurately predict the band gap and electronic structure dispersion. The charge transfer between the transition metal atoms and the monolayer surface is quantitatively analysed by Bader charge calculations[31, 32]. The thermal stability of the adsorbed monolayer was assessed through ab-initio molecular dynamics (AIMD) simulations in an NVT ensemble for 10 ps with a time step of 0.2 fs, where the temperature was maintained by Nos´e–Hoover thermostat[33].

The adsorption energy of the pristine and transition metal-decorated aza-triphenylene monolayer was computed using the relation

$$E_{ads} = E_{surface+H} - E_{surface} - \frac{1}{2}E_{H_2} \quad (1)$$

Where $E_{surface+H}$, $E_{surface}$ and $E_{H_2}$ indicates the total energy of an aza-COF surface with hydrogen, an aza-COF surface without hydrogen and a free $H_2$ molecule, respectively. The more negative the value of the adsorption energy, the better the thermodynamic stability the system can have.

## Results and Discussions

### Crystal Structure and Electronic Properties

Figure 1 (a,b) illustrates the single unit cell of an aza-triphenylene monolayer forming a covalent organic framework (AZA-COF). The top view of the unit cell (Figure 1a) indicates the distribution of carbon, nitrogen and hydrogen atoms forming the monolayer through interconnected benzene and pyrazine rings. There are two types of benzene rings. The central benzene ring contains two hydrogen atoms at the para position, and the ortho and meta carbon atoms forming the C-C bond are bridged by the neighbouring pyrazine rings. The other type of benzene ring is bonded to three pyrazine rings connected to the three alternate C-C sides of the benzene ring. The AZA-COF forms a porous material where the pore size is calculated to be around 14.43 Å, consistent with the experimentally measured value of 14.5 Å[34]. The side view of the unit cell indicates that the monolayer possesses a planar structure (Figure 1b). The calculated lattice constant, after geometrical optimisation, is found to be 16.5 Å, which closely aligns with the reported value of 16.47 Å[35]. One unit cell of aza-triphenylene monolayer contains a total of 48 atoms, which are made up of 30 carbon, 12 nitrogen, and 6 hydrogen atoms, respectively. The calculated bond length for the C-C bond is around 1.40 Å, which varies within 0.5 Å for the central and other benzene rings. Similarly, the C-N bond length within the pyrazine rings is calculated to be around 1.32 Å, which agrees quite well with previous reports[36]. The distribution of electrons on the surface of the two-dimensional aza-triphenylene monolayer is analysed by calculating the electron localization function (ELF) (Figure 1c). The ELF is a mapping for the electron density distribution which varies between zero to one (see the scale bar in Figure 1c). The ELF values near zero (blue regions) and one (red regions) indicate delocalization and substantial electron concentration, respectively. The ELF with intermediate values (between zero and one), as indicated in the green region in Figure 1c, represents the electron gas-like state. From Figure 1c, it is clear that electrons are largely accumulated near the nitrogen and hydrogen atoms, respectively. In contrast, electrons are evenly distributed over the C-C and C-N bond which represents the conjugated π-bond characteristic of the monolayer. Overall, the ELF calculations indicate the formation of a stable covalent network ensuring the structural integrity of the sample. The calculated electronic band structure of the pristine aza-triphenylene, using the PBE-GGA functional, is shown in Figure 1d. The occurrence of the valence band maximum (VBM) and conduction band minimum (CBM) at the identical 'Γ' point signifies the direct band gap characteristic of the specimen. The magnitude of the electronic band gap gap is around 1.1 eV, which is in agreement with previous DFT predictions[35-38]. The projected density of states calculations (Figure 1e) indicate that the VBM is contributed by the '2p' states of carbon, whereas the CBM has a major contribution from the nitrogen '2p' states. To precisely determine the band gap, we have performed the band structure calculations using HSE-06 hybrid functional, which estimated an energy gap around **1.3 eV,** corroborating earlier results[39] (see Figure 1f).

**Gibbs' free energy calculation for HER**

Hydrogen production via electrochemical water splitting consists of two fundamental reaction stages. The initial step involves the adsorption of atomic hydrogen at the catalytic site, and the second step is the evolution of molecular hydrogen. To analyse the HER performance of the pristine aza-triphenylene monolayer, the Gibbs' free energy change calculation was performed during the adsorption of hydrogen.

$$H^{+} + e^{-} \xrightarrow{\Delta G_H} H^{*} \quad (2)$$

Here H* indicates the aza-COF adsorbed with hydrogen. The change in Gibbs' free energy for the adsorption of hydrogen was assessed using the relation

$$\Delta G_{HER} = \Delta E_{ads} + \Delta E_{ZPE} - T\Delta S_H + \Delta G_{pH} + \Delta G_U \quad (3)$$

Where the first term $\Delta E_{ads}$ indicates the adsorption energy of hydrogen, $\Delta E_{ZPE}$ and $T\Delta S_H$ indicates the zero-point energy and entropy change between the adsorbed hydrogen and $H_2$ gaseous phase. $\Delta G_{pH}$ and $\Delta G_U$ represents the pH dependency and external potential bias (U=0) respectively. T is the temperature in absolute scale (Here we take T= 298.15 K). The entropy of the molecular hydrogen was obtained from the NIST database[40]. The presence of a strong adsorption energy makes the hydrogen desorption difficult. Similarly, a weak adsorption energy makes the attraction difficult between the hydrogen and the catalyst surface. Therefore, for an efficient catalyst for HER, the adsorption energy ought to be neither excessively strong nor excessively weak. According to Sabatier principle, for optical HER performance with maximum reaction rate, the Gibbs' free energy change must approach to zero.

The Gibb's free energy change for the HER in pristine aza-triphenylene monolayer is shown in Figure 2a. After incorporating the zero-point energy and correction in entropy, the $\Delta G_{HER}$ becomes around -0.78 eV. This large negative value indicates that the hydrogen adsorption is thermodynamically too strong, which is not useful for HER. Therefore, the pristine sample is expected to exhibit limited HER activity.

In the next step, to improve the HER catalytic activity, the aza-triphenylene monolayer was decorated with a wide range of transition metals (Cr, Mn, Mo, Pd, Pt, Rh, Ru, Ti, V, Y, Zr, Fe, Co), as single-atom catalysts. For each of these transition metals, different suitable adsorption sites were selected for ionic relaxation, and the adsorption site exhibiting the lowest total energy was regarded as the ground state for further computations. It was observed that, out of all the possible adsorption sites considered, all the transition metals exhibit a preference for adsorption adjacent to the nitrogen atom of the pyrazine rings, forming chemical bonds with the nitrogen atoms. Using the relations 1 & 2, the adsorption energy and Gibbs' free energy change were calculated for all the transition metal adsorbed structures. The detailed calculation of the Gibbs' free energy change is presented in Table S1 in the Supporting Information file. It was observed that, among all the traisition metal considered, Ru, Co and Y adsorbed (hereafter named as Ru@aza, Co@aza and Y@aza) aza-triphenylene monolayer exhibited $\Delta G_{HER}$ around 0.20 eV, 0.05 eV and 0.06 eV respectively (Figure 2b). The calculated values of HER overpotentials of the present structure are consistent with the HER overpotential of other 2D materials reported in the literature (see Table S2 in the Supporting Information file). As these values are extremely low and very close to zero, Ru@aza, Co@aza and Y@aza are highly promising electrocatalysts for carrying out the HER. The schematic representation of Co-, Ru- and Y-adsorbed aza-triphenylene monolayer, after geometrical optimization is presented in Figure S1 of the Supporting Information file. For a deep understanding of the inherent electron rate in the HER at the equilibrium potential, we have computed the theoretical exchange current density ($i_0$), using Norskov's assumption[41] under the standard conditions (T= 298.15 K, $pH = 0$), which is given as follows.

For an exothermic reaction ($\Delta G_{HER}$< 0),

$$i_0 = \frac{-ek_0}{1 + \exp\left(-\Delta G_{HER}/K_B T\right)} \tag{4}$$

For an endothermic reaction ($\Delta G_{HER}$ > 0),

$$i_0 = \frac{-ek_0 \exp\left(-\Delta G_{HER}/K_B T\right)}{1 + \exp\left(-\Delta G_{HER}/K_B T\right)} \tag{5}$$

Where $K_B$, $k_0$ denote the Boltzmann constant and rate constant, respectively. The exchange current density ($i_0$) is a parameter that estimates the intrinsic activity of a catalyst for efficient HER performance. The analysis is graphically presented as a volcano plot, which is a graph between the theoretical exchange current density and the Gibbs' free energy change ($\Delta G_{HER}$), as illustrated in Figure 2c. For the optimal HER activity, the exchange current density should be maximum as $\Delta G_{HER}$ approaches to zero. From the volcano plot in Figure 2c, it is clear that Co-, Ru and Y-adsorbed aza-triphenylene exhibit the maximum exchange current density with an extremely low value of $\Delta G_{HER}$. This observation is consistent with the HER overpotential plot which further confirms that Co, Ru and Y adsorbed aza-triphenylene can be the best promising candidate for efficient HER performance.

**Ab Initio Molecular Dynamics (AIMD) Stability Analysis**

To ensure the thermal stability of the pristine aza-triphenylene monolayer and the transition-metal-decorated systems, ab initio molecular dynamics (AIMD) simulations were executed for 10 ps. The corresponding energy fluctuation profiles for the pristine, Co@aza, Ru@aza, and

Y@aza monolayers are shown in Fig. 3 (a-d). As observed, all systems demonstrate only small oscillations in total energy throughout the simulation time without any abrupt energy drift or structural collapse, indicating good thermodynamic stability under finite-temperature (T = 300 K) conditions. The pristine aza-triphenylene monolayer maintains a nearly small energy fluctuation around the mean value (-389 eV), confirming its intrinsic structural robustness. Similarly, the Co-, Ru-, and Y-decorated systems exhibit stable energy fluctuations throughout the simulation period, suggesting that the adsorbed metal atoms remain firmly anchored to the monolayer surface without aggregation or detachment. Among the decorated systems, Ru@aza and Y@aza exhibit slightly larger fluctuations than the pristine structure, which can be attributed to enhanced local atomic vibrations resulting from the incorporation of metal atoms. Nevertheless, the absence of significant energy variation demonstrates that all metal-decorated configurations possess favourable thermal stability. In addition, the AIMD trajectories demonstrate uniform fluctuations of energy around the corresponding mean values, lacking any substantial energy drift. The maximum discrepancies from the mean energy are 0.0419, 0.0202, 0.0208, and 0.0189 eV/atom for pristine aza, Co@aza, Ru@aza, and Y@aza, respectively, These findings confirm that the proposed pristine and metal-functionalized aza-triphenylene monolayers are thermodynamically stable and suitable candidates for HER applications under operational conditions.

## COHP Analysis

To obtain a profound understanding of the chemical interactions between the adsorbed metal atoms and the aza-triphenylene framework, crystal orbital Hamilton population (COHP) analysis was conducted for the M–N and M–H interactions, where M = Co, Ru, and Y, as shown in Figure 4. The COHP curves present an energy-discriminated analysis of the bonding and antibonding factors, while the corresponding integrated COHP (ICOHP) values evaluate the overall strength of interaction. In the present convention, negative COHP contributions correspond to bonding states, whereas positive contributions represent antibonding states. Accordingly, a more negative ICOHP value indicates a more robust net bonding relationship.

For the metal–nitrogen interactions, the calculated ICOHP values are −2.42, −2.05, and −3.59 eV for Co–N, Ru–N, and Y–N, respectively, in Figure 4 (a–c). The Y–N interaction, therefore, exhibits the largest magnitude of ICOHP, indicating the strongest net bonding among the three metal–N interactions. The pronounced bonding contributions associated with the occupied states below the Fermi level further support the strong interaction between Y and the neighbouring N atoms. In comparison, the Co–N interaction is moderately stronger than Ru–N, as reflected by its more negative ICOHP value. These results demonstrate that the interaction between the metal dopant and the aza framework strongly relies on the chemical characteristics of the incorporated metal.

The corresponding metal–hydrogen interactions exhibit a different sequence of bonding strengths, Fig. 4(d–f). The ICOHP values obtained for Co–H, Ru–H, and Y–H are −2.26, −2.94, and −2.03 eV, respectively. Among the investigated systems, Ru–H displays the most negative ICOHP value, illustrating the strongest net Ru–H interaction, whereas Y–H exhibits the weakest interaction based on ICOHP magnitude. The Co–H interaction lies between these two cases. The COHP distributions also reveal substantial bonding contributions below the Fermi level for all three metal–H pairs, accompanied by antibonding contributions predominantly above the Fermi level. This indicates that the occupied electronic states contribute favourably to metal–hydrogen bonding, while the corresponding antibonding states are largely unoccupied.

Interestingly, the bonding-strength trends of the metal–N and metal–H interactions are not identical. While Y forms the strongest interaction with the N atoms of the AZA framework, Ru exhibits the strongest interaction with H. This difference suggests that the strength of hydrogen binding cannot be inferred solely from the metal–framework interaction. Rather, it reflects the electronic characteristics of the metal centre and its orbital interaction with the adsorbed hydrogen. The relatively strong Ru–H interaction is consistent with enhanced orbital interaction between Ru and H, whereas the comparatively weaker Y–H interaction suggests a less strongly bound hydrogen species. Such a balance between metal–substrate stabilisation and metal–hydrogen interaction is particularly important when evaluating transition-metal-decorated two-dimensional materials for HER applications.

**Electronic Structure and Projected Density of States**

In this section, we analyse the electronic structure of aza-triphenylene monolayer decorated with Co, Ru and Y. The spin-polarised band structure for these adsorbed structures adopting the PBE-GGA functional is given in Figure 5 (a-c). The calculations predict a very narrow band gap state for all the configurations where the CBM and VBM are formed by the spin-up and spin down channel respectively. The magnitude of the energy gap for Co@aza, Ru@aza, and Y@aza structures is estimated to be around 0.07 eV, 0.15 eV and 0.09 eV, respectively. Since the PBE-GGA functional is widely known for its underestimation of the band gap due to derivative discontinuity and self-interaction error, we have performed the band structure calculations using the HSE-06 hybrid functional as well, which are illustrated in Figure 5 (d-f). The hybrid calculation exhibits a comparatively larger energy gap for the decorated structure. For the Co@aza structure, the energy gap is around 0.35 eV, where the VBM and CBM occur at the 'M' and 'Γ' points, respectively (Figure 5d). In contrast, the Ru-adsorbed structure (Ru@aza) exhibits a larger band gap around 0.98 eV, where the VBM and CBM form an indirect band gap state (Figure 5e). It is interesting to note that for Ru@aza structure, an almost flat band was observed at the 'Γ' point at around 1eV above the Fermi level which can be attributed to its low effective mass. On the other hand, for the Y-decorated structure (Y@aza), the structure exhibits a large band gap around 0.87 eV, where both the CBM and VBM occur along the $K - \Gamma$ direction (Figure 5f).

To ascertain the role of various orbitals in the formation of the CBM and VBM, calculations of the projected density of states (PDOS) were conducted for the Co, Ru and Y adsorbed aza-triphenylene monolayer (see Figure 6). From the spin-polarized PDOS plot in Co@aza structure, it is evident that the major contribution close to the Fermi level, forming the VBM and CBM, comes from the '2p' orbitals of nitrogen and '3d' orbitals of Co, forming a narrow band gap state (Figure 6a). The carbon '2p' states are dispersed at a larger energy with reference to the Fermi level. In contrast, for Ru adsorbed aza-triphenylene (Ru@aza) the VBM and CBM are well separated in energy, where the VBM has a major population from Ru '4d' orbitals and the CBM is primarily dominated by the '2p' orbitals of N (see Figure 6b). Interestingly, the carbon '2p' orbitals are found to be very close to the VBM. On the other hand, for the Y-adsorbed aza-triphenylene monolayer, the VBM is dominated by nitrogen '2p' states and the CBM has a nearly equal population from '2p' states of carbon and nitrogen, respectively (see Figure 6c). The yttrium exhibits a non-magnetic nature with equal contribution from spin-up and spin-down channels and mostly dispersed in the conduction band region.

**Deformation Potential Analysis**

The deformation-potential behaviour of the pristine aza-triphenylene monolayer and metal-decorated systems was investigated under biaxial strain ranging from -1% to +1%, and the corresponding shifts in the valence band maximum (VBM) and conduction band minimum (CBM) are shown in Fig. S2 (a-d) in the Supporting Information file. Such an approach, in which the evolution of the band-edge energies with applied strain is used to evaluate the deformation-potential response, has been widely adopted in first-principles investigations of two-dimensional materials[42, 43]. For the pristine aza-triphenylene monolayer, both the VBM and CBM exhibit very small linear variation with the applied strain range, indicating excellent electronic stability and weak sensitivity to small lattice deformations (Figure S2(a) in the Supporting Information file). Such behaviour suggests that the pristine system possesses a mechanically robust electronic structure that is minimally perturbed by external strain.

In contrast, the metal-decorated systems exhibit comparatively pronounced shifts in the band-edge positions as strain increases. For the Co@aza monolayer, both the VBM and CBM shift gradually toward lower energies as the strain changes from compressive to tensile mode (Figure S2(b) in the Supporting Information file). This continuous variation reflects enhanced coupling between the Co atom and the aza framework under lattice deformation, thereby effectively modulating the system's electronic properties. An analogous tendency is similarly recognized for the Ru@aza monolayer, where the band-edge potentials decrease almost linearly with increasing strain, indicating strain-dependent electronic tunability induced by Ru incorporation (Figure S2(c) in the Supporting Information file). A comparable trend is evident for the Y@aza monolayer, where both the VBM and CBM potentials decrease progressively as strain increases, from compressive to tensile conditions (Figure S2(d) in the Supporting Information file). The nearly linear variation of the band-edge positions indicates that the electronic structure of Y@aza is systematically influenced by lattice deformation. This strain-dependent response can be ascribed to the interplay between the Y atom and the aza-triphenylene framework, which modulates the electronic states under applied biaxial strain. Thus, Y decoration, like the other metal-decorated systems, provides an effective means of tuning the electronic properties of the aza-triphenylene monolayer through mechanical strain. Overall, the deformation potential analysis demonstrates that the decorated systems possess tunable electronic characteristics under external strain, which is beneficial for optimising charge-transfer behaviour and catalytic activity.

**Calculation of carrier mobility**

The performance of a material for efficient hydrogen evolution reaction, photovoltaic, and photocatalytic applications is governed by an important parameter, “carrier mobility”, which plays the role of a crucial descriptor like energy gap and effective separation of charge carriers.

In the current study, the carrier mobility ($\mu_{2D}$) of 2D pristine and transition metal-adsorbed aza-triphenylene monolayer is calculated by implementing the effective mass approximation and the deformation potential theory established by Bardeen and Shockley[44]. This approach primarily considers carrier scattering by long-wavelength acoustic phonons and provides an upper limit to the carrier mobility[45] and takes into account the anisotropic nature of carrier transport. The carrier mobility for a 2D material can be expressed as

$$\mu_{2D} = \frac{e\hbar^3 C_{2D}}{K_\beta T\, m^{*2} {E_d}^2} \tag{6}$$

Where, $e$, $\hbar$, $K_\beta$ indicates the electronic charge, reduced Planck constant and Boltzmann constant, respectively. The term $C_{2D}$ represents the stretching modulus which can be calculated by using the expression

$$C_{2D} = 2\,\frac{\left[\frac{\partial^2 E_{total}}{\partial \delta^2}\right]}{A} \tag{7}$$

Where, $\delta$, $E_{total}$ and $A$ represents the applied strain, total energy and area of the equilibrium unit cell, respectively. The effective mass can be determined from the band structure using the expression

$$m^* = \frac{\hbar^2}{\frac{\partial^2 E}{\partial K^2}} \tag{8}$$

Where K and E denote the wave vector and the corresponding energy at the particular wave vector, respectively. The effective mass is calculated in the vicinity of the VBM and CBM along the high-symmetry path M-Γ direction. The deformation potential estimates the scattering of charge carriers along the periodic direction by acoustic phonons. The expression for calculating the deformation potential is given by

$$E_d = \frac{dE_{edge}}{d\delta} \tag{9}$$

Here $E_{edge}$ is the energy of the VBM and CBM under biaxial strain. Using all the above expressions, the stretching modulus, deformation potential, effective mass of electrons and holes, deformation potential and the carrier mobility for the electrons and holes are given in Table 1. The stretching modulus is calculated using equation 7, by the strain-dependent total energy variation curve and taking a polynomial fit (see Figure S3 in the Supporting Information file). The deformation potential is calculated using equation 9 by analysing the alterations of band edges as a function of strain and taking the slope with a linear fit (see Figure S2 in the Supporting Information file). Using the above relations, the calculated values of carrier mobilities for pristine and adsorbed monolayers is given in Table 1. For pristine aza-triphenylene, the carrier mobilities are estimated to be $5.20 \times 10^{3}\ cm^{2}V^{-1}S^{-1}$ (for electrons) and $8.32 \times 10^{-4}\ cm^{2}V^{-1}S^{-1}$ (for holes) respectively. However, after decorating with Co and Y, the electron mobility was reduced to $2.23 \times 10^{1}\ cm^{2}V^{-1}S^{-1}$ and $3.65 \times 10^{2}\ cm^{2}V^{-1}S^{-1}$ respectively, which is less than almost two and one order of magnitude as compared to the pristine monolayer. In contrast, the hole mobility of the Co and Y-adsorbed aza-triphenylene exhibits a significant increase to $2.93 \times 10^{2}\ cm^{2}V^{-1}S^{-1}$ and $1.23 \times 10^{3}\ cm^{2}V^{-1}S^{-1}$ respectively. It is interesting to note that the Ru-decorated aza-triphenylene exhibits a substantial enhancement in both electron and hole mobility to $2.77 \times 10^{6}\ cm^{2}V^{-1}S^{-1}$ and $2.97 \times 10^{6}\ cm^{2}V^{-1}S^{-1}$ respectively. Such a significant enhancement in carrier mobility of Ru@aza can originate from the extremely low effective mass of the electrons and holes, respectively.

In this section, we concentrate on the reaction kinetics of the hydrogen evolution reaction using the climbing image nudged elastic band method. The mechanism can be studied under acidic and alkaline media. However, due to computational limitations, we restricted our analysis to acidic medium only. The reaction begins with the adsorption of a single hydrogen atom on the surface of the catalyst, which is known as the Volmer reaction. The final step is the desorption of molecular hydrogen, which can occur in one of the two possible reaction mechanisms. The first one is known as the Heyrovsky reaction, where an adsorbed hydrogen atom on the catalyst surface reacts with another hydrogen atom to form a hydrogen molecule. The subsequent process is referred to as the Tafel reaction, where two hydrogen atoms adsorbed at two different locations on the catalyst surface interact with each other and form a hydrogen molecule. In the presence of an acidic medium, the electrochemical HER can happen in either of the following two competing strategies, known as the Volmer-Heyrovsky reaction and Volmer-Tafel reaction respectively. These reaction schemes can be represented by the following equations.

$$H^{+} + e^{-} \rightarrow H^{*}\ (Volmer) \qquad (10)$$

$$H^{*} + H^{+} + e^{-} \rightarrow H_{2}\ (Heyrovsky) \qquad (11)$$

$$H^* + H^* \rightarrow H^* \ (Tafel) \tag{12}$$

Here $H^*$ indicates the hydrogen atom adsorbed on the surface of the catalyst. Out of the two possible reaction schemes, one mechanism dominates depending on the nature of the catalyst and determines the rate of the reaction. During the adsorption and desorption analysis, two parameters play an important role in determining the transition path connecting the initial and final state of the reaction. These parameters are known as the activation energy ($E_a$) and reaction energy ($\Delta E$) respectively. The former one is the difference in energy between the initial state and the transition state, whereas the latter is the difference in energy between the initial and final state of the reaction. The transition path during the adsorption of atomic hydrogen and desorption of molecular hydrogen are analysed by adopting the climbing image nudged elastic band method.

In the present work, out of the wide range of transition metal considered, Co adsorbed aza-teiphenylene monolayer (Co-aza) exhibits the lowest Gibbs' free energy change around 0.05 eV. Therefore, we have performed the HER mechanism for Co-aza system only.

The minimum energy path during Volmer, Heyrovsky and Tafel reactions in Co-aza system is illustrated in Figure 7. For the Volmer reaction, a single hydrogen atom was placed across all feasible adsorption sites and an ionic relax calculation was executed. The hydrogen atom was found to have its lowest energy when adsorbed on the Co atom which is known as the active site. For the initial structure in Volmer scheme, the initial separation between the Co and hydrogen atom was set at 5 Å which reduces to 1.54 Å after forming a bond with the Co atom in the final state. The transition from initial to final state occurs through several intermediate metastable states with an activation energy barrier of 1.61 eV which occurs due to the change in Co-N bond length from 1.91 Å to 1.84Å (Figure 7a). A comparable energy barrier of 1.23 eV was observed for the Volmer reaction in Biphenylene nanoribbons[10].

In the next step, we analyse the HER mechanism under Heyrovsky reaction scheme (Figure 7b). Here the adsorbed hydrogen was allowed to react with another hydrogen atom with an initial separation of 3 Å. In the final step a hydrogen molecule is formed with H-H bond length around 0.75 Å. The adsorption of atomic hydrogen and release of molecular hydrogen was analysed through the transition state profile calculated using NEB calculations. As the second hydrogen approached the adsorbed hydrogen their separation decreases to 2.57 Å. In the second intermediate step the transition state occupies a higher energy level and a large energy barrier around 2.29 eV was observed due to the dissociation of the Co-H bond. As the reaction proceeds, the two hydrogen atoms combine to form hydrogen molecules, which is accompanied by a sudden reduction in energy as observed in the third intermediate step. For the fourth, fifth intermediate configurations, a very small change in energy was observed due to a change in orientation of the hydrogen molecule. In the final structure the molecular hydrogen was found around 4.3 eV above the monolayer surface.

In the final step, the reaction mechanism under the Tafel reaction scheme was analysed. In this case, the two hydrogen atoms were bonded to the Co atom and N atoms of the pyrazine ring respectively. Initially, the bond length between the Co-H and N-H bonds was around 1.55 Å and 1.02 Å, respectively. The transition path describing the evolution of the hydrogen molecule calculated using the NEB method, is given in Figure 7c. The reaction begins with the interaction between the two bonded hydrogen atoms. In the first intermediate state the structure achieves a higher energy due to change in orientation of Co-H bond. In the second intermediate state, both the Co-H and N-H bonds start interacting with each other, leading to a higher energy state. The third transition state involves dissociation of the Co-H bond, which needs a large amount of energy. For the fourth intermediate image, the second hydrogen dissociates from the N-H bond, leading to an extremely large energy barrier around 5.88 eV. As the reaction proceeds, the two dissociated hydrogen atoms combine to form a hydrogen molecule, which is reflected in a decrease in the system's energy, as shown in the fifth intermediate image. In the final configuration, the hydrogen molecule slightly changes its orientation and gets stabilised without exhibiting a substantial change in energy as compared to its previous state. Therefore, the Tafel reaction seems to be very unlikely due to the presence of a very large energy barrier associated with the dissociation of two adsorbed hydrogen atoms.

Since the Tafel reaction exhibits a very large energy barrier of 5.88 eV, the Volmer-Heyrovsky reaction is the dominant mechanism for the HER process in a Co-adsorbed aza-triphenylene monolayer, with the Heyrovsky reaction serving as the rate-determining step.

**d-band centre analysis**

In order to have a deep understanding of the source of catalytic activity, the d-band centre calculation was performed for the three adsorbed structures (Co@aza, Ru@aza and Y@aza) for the catalytically active sites. The d-band centre can provide deep insight for the HER activity and can be calculated from the following expression.

$$\epsilon_d = \frac{\int_{-\infty}^{\infty} E\,\rho(E)\,dE}{\int_{-\infty}^{\infty} \rho(E)\,dE} \tag{13}$$

Where $\rho(E)$ is the projected density of states for the d-orbitals at a particular energy E. The analysis was executed by performing the integration in the valence band region for the occupied state in the energy range -6 eV to zero. The catalytic activity is strongly influenced by the electronic states available close to the Fermi level. If the d-band centre value is low with reference to the Fermi level, this indicates a robust interaction between the catalyst and the adsorbed hydrogen atom, which results in a small value of the $\Delta G_{HER}$. Similarly, if the d- band centre value is more positive then the $\Delta G_{HER}$ also gives a positive value. The calculated d-band centre value for the Co, Ru and Y adsorbed aza-triphenylene monolayer is given in the Table 1, which were estimated to be -1.452 eV, -1.826 eV and -3.162 eV, respectively. These values are lying far from the Fermi level leading to a small value of $\Delta G_{HER}$.

The catalytic activity of a 2D material is largely affected by the transfer of electronic charge between the catalyst and the hydrogen atoms. In order to analyse the charge transfer mechanism, the charge density difference was computed by calculating the charge density of the adsorbed monolayer and then subtracting from it the charge densities of the substrate and the hydrogen. The relation for calculating the charge density is given by

$$\Delta\rho = \rho_{aza-COF+TM+H} - \rho_{aza-COF+TM} - \rho_H \tag{14}$$

Where $\Delta\rho$ represent the charge density difference, $\rho_{aza-COF+TM+H}$, $\rho_{aza-COF+TM}$ and $\rho_H$ represent the charge density of the transition metal adsorbed aza-triphenylene with hydrogen, the charge density of the transition metal adsorbed aza-triphenylene without hydrogen and the charge density of atomic hydrogen, respectively. The calculated charge density difference for the Co-, Ru-, and Y-adsorbed monolayer is shown in Figure 8. In all the Figures, the cyan colour denotes a depletion of charge and the yellow colour denotes an accumulation of electronic charges respectively. For the Co@aza and Ru@aza structures (Figure 8 a,b), the cyan colour appears near the transition metal catalyst and yellow colour appears at the hydrogen atom. Therefore, there is a transfer of electronic charge from Co/Ru to the hydrogen. In contrast, for the Y@aza structure (Figure 8c), the cyan colour appears at the hydrogen and yellow colour appears at the Y atom, indicating a charge loss at the hydrogen and a charge gain at the Y atom. This behaviour is in contrast to the Ru@az and Co@aza configurations. To further analyse the charge transfer quantitatively, the Bader charge calculations were performed. For the Co@aza structure, a negative charge of around -0.29 e, appears at the hydrogen and a positive charge (+0.725 e) appears at the Co atom, respectively (Figure 8d). Similarly, for the Ru@aza structure, a negative charge around -0.249 e occurs at the hydrogen atom and a positive charge of +0.714 e develops on the Ru atom (Figure 8e). For both Co@aza, and Ru@aza, charge transfer occurs from the transition metal (Co/Ru) to the adsorbed hydrogen. On the other hand, for the Y@aza structure, a very negligible charge of -0.001 e appears on the hydrogen atom and a positive charge of +1.587e develops on the Y atom (Figure 8f). This observation indicates that the adsorbed hydrogen remains almost neutral and the interaction between Y and H is relatively weak as compared to the Co-H and Ru-H interaction.

**Conclusion**

In conclusion, the HER activity of the aza-triphenylene monolayer is investigated for the pristine and transition metal adsorbed configurations using first-principles calculations. The pristine aza-triphenylene exhibits a direct band gap of 1.3 eV and exhibits a large negative HER Gibbs' free energy of -0.78 eV. After decorating with a large no of single transition metal on the surface, the Gibbs' free energy change becomes extremely low, specifically 0.2 eV, 0.06 eV and 0.05 eV for Ru-, Y-, and Co-adsorbed structures, respectively. The Volcano plot also confirms this observation where the peak of the curve is maximum for Ru, Co and Y where, $\Delta G_{HER}$ becomes close to zero. The electronic band structure exhibits a reduced band gap for the adsorbed structures as compared to the pristine configurations. The calculated charge density difference, along with Bader charge analysis, confirms a charge transfer between the catalyst and adsorbed hydrogen, which is responsible for the enhanced HER activity. The d-band centre for the Co, Ru and Y adsorbed structures becomes negative and close to the Fermi level, indicating a strong interaction between the catalyst and the adsorbed hydrogen atom. The carrier mobility after transition metal adsorption shows a substantial improvement and becomes very high for the Ru-adsorbed monolayer due to having a very low effective mass for electrons and holes. The calculated reaction path using NEB method exhibits an energy barrier of 1.61 eV and 2.29 eV for the Volmer and Heyrovsky reactions, whereas a large activation energy barrier around 5.88 eV is observed for the Tafel reaction. This makes the Volmer-Heyrovsky reaction a possible reaction pathway for hydrogen evolution with the Heyrovsky reaction as the rate-determining step, whereas the Tafel reaction becomes energetically unfavourable due to the presence of a very large energy barrier. The present work predicts theoretically the possibility of enhancing HER activity through transition metal adsorption and paves the way for experimental design of artificial 2D materials for efficient hydrogen evolution reaction for future energy storage applications.

**ACKNOWLEDGMENTS**

The authors gratefully acknowledge the use of high-performance computing infrastructure made available by the National Institute of Science Education and Research (NISER) and the Bhabha Atomic Research Centre (BARC).

**DATA AVAILABILITY STATEMENT**

The data supporting the findings of this study are available from the corresponding author upon reasonable request.

**FUNDING**

The authors declare that no external funding was received for this research.

**CONFLICT OF INTEREST**

The authors declare that they have no conflicts of interest.

**SUPPORTING INFORMATION**

Additional figures supporting the findings of this study are provided in the Supporting Information file.

| System | $E_d(h)$ (eV) | $E_d(e)$ (eV) | $m_h^*$ ($\times m_0$) | $m_e^*$ ($\times m_0$) | $C_{2D}$ ($\times 10^{21}$ $eV\,m^{-1}$) | $\mu_{2D}(e)$ ($cm^2V^{-1}S^{-1}$) | $\mu_{2D}(h)$ ($cm^2V^{-1}S^{-1}$) | d-band centre (eV) |
|---|---|---|---|---|---|---|---|---|
| Pristine | -0.053 | -0.041 | -0.299 | 578.378 | 2.292 | $5.20 \times 10^3$ | $8.32 \times 10^{-4}$ | |
| Co@aza | -0.039 | -0.048 | -3.949 | -1.341 | 2.352 | $2.23 \times 10^1$ | $2.93 \times 10^2$ | -1.452 |
| Ru@aza | -0.058 | -0.060 | -0.009 | -0.009 | 2.369 | $2.77 \times 10^6$ | $2.97 \times 10^6$ | -1.826 |
| Y@aza | -0.072 | -0.068 | 0.678 | -0.349 | 2.272 | $3.65 \times 10^2$ | $1.23 \times 10^3$ | -3.162 |

**TABLE 1.** Calculated values of deformation potential (E*d*), effective mass (m*), stretching modulus ($C_{2D}$), and carrier mobility ($\mu_{2D}$) for electrons (e) and holes (h). $m_0$ refers to the rest mass of the electron.

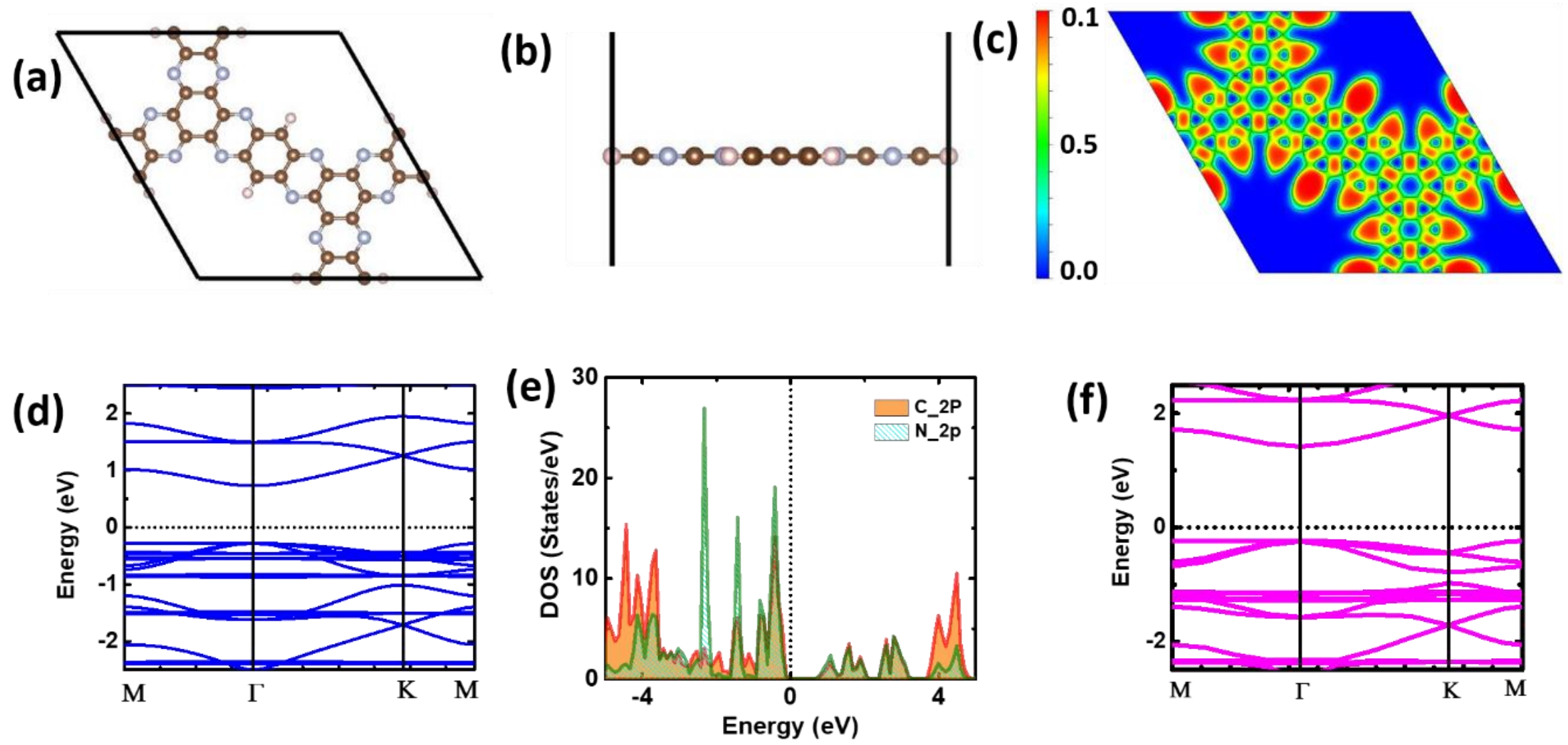

Figure 1: Schematic illustration of (a) top and (b) side view, (c) electron localisation function, (d) band structure and (e) projected density of states using the PBE-GGA functional, (f) HSE-06 band structure of pristine aza-triphenylene monolayer. In the band structure and PDOS plot, the Fermi level was set to zero as indicated by the dotted horizontal and vertical lines, respectively.

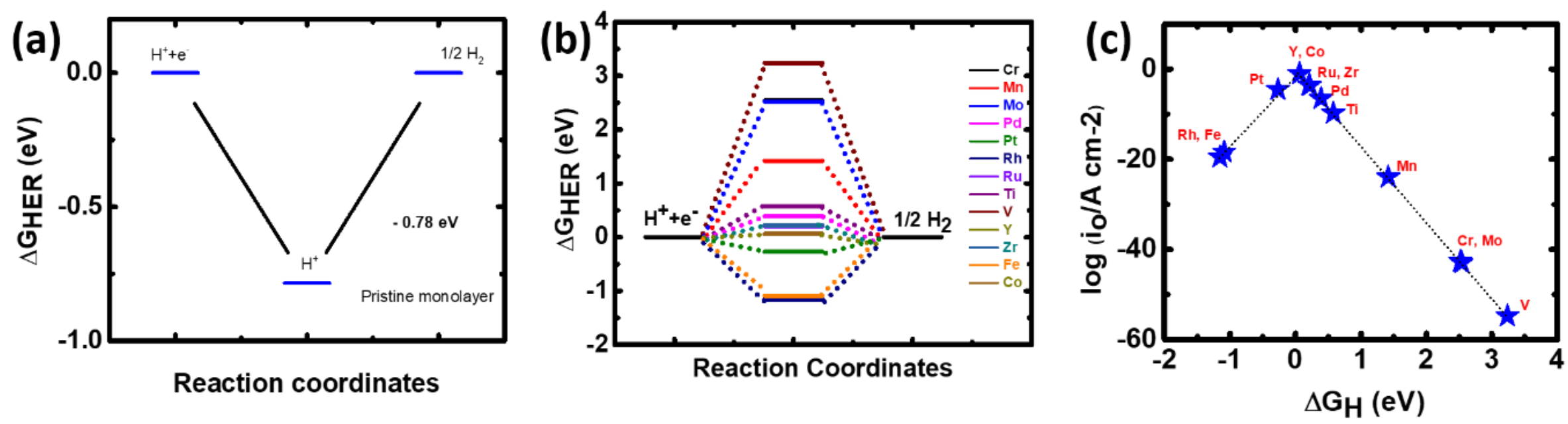

Figure 2: Calculated Gibbs' free energy change for hydrogen evolution reaction ($\Delta G_{HER}$) in (a) pristine and (b) transition metal decorated aza-triphenylene monolayer. (c) The Volcano plot showing the variation of theoretical exchange current density vs $\Delta G_{HER}$ for the different adsorbed transition metals.

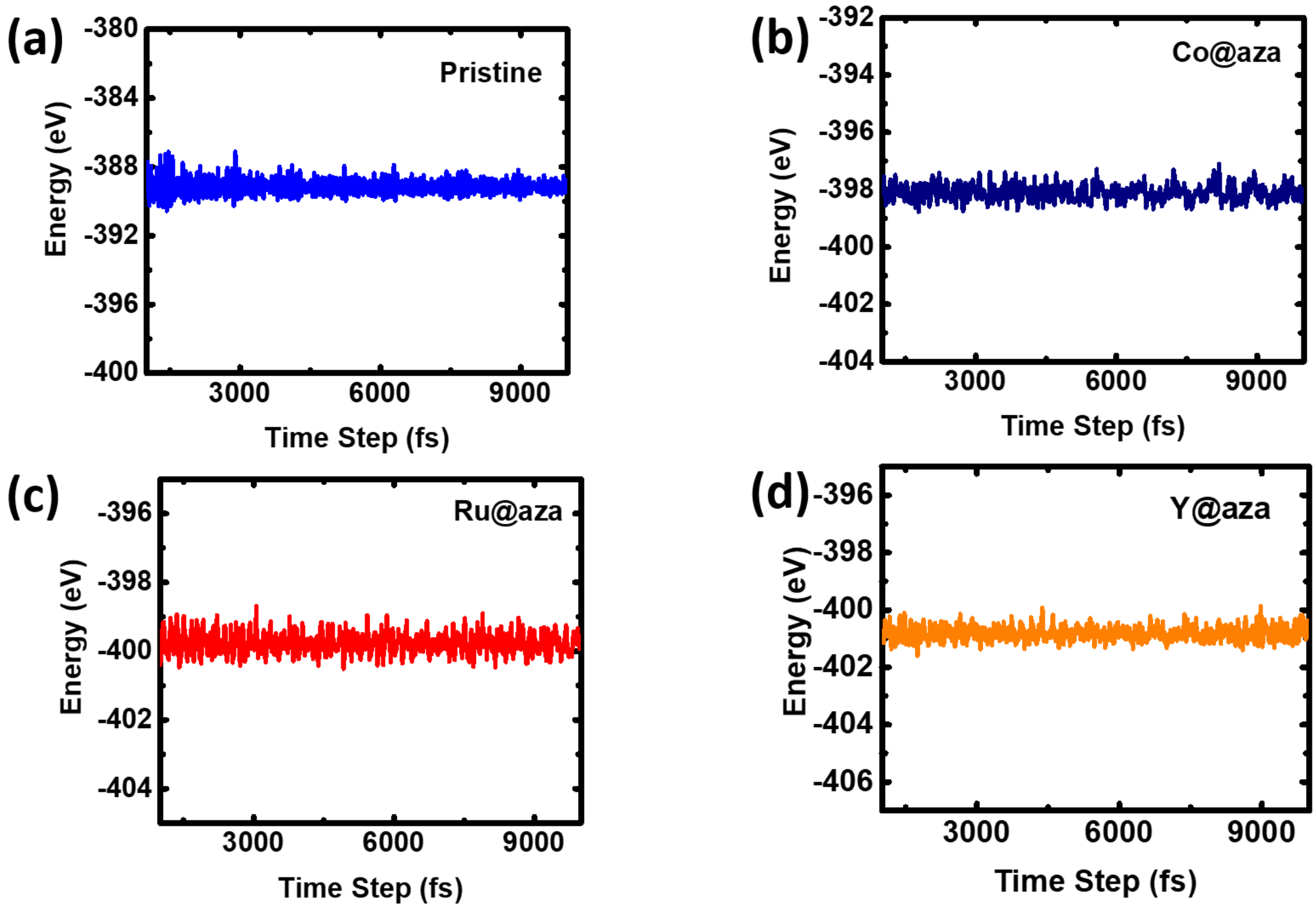

Figure 3: Ab-initio molecular dynamics simulation (AIMD) showing the energy fluctuation as a function of simulation time in (a) pristine, (b) Co adsorbed, (c) Ru adsorbed, (d) Y adsorbed aza-triphenylene monolayer.

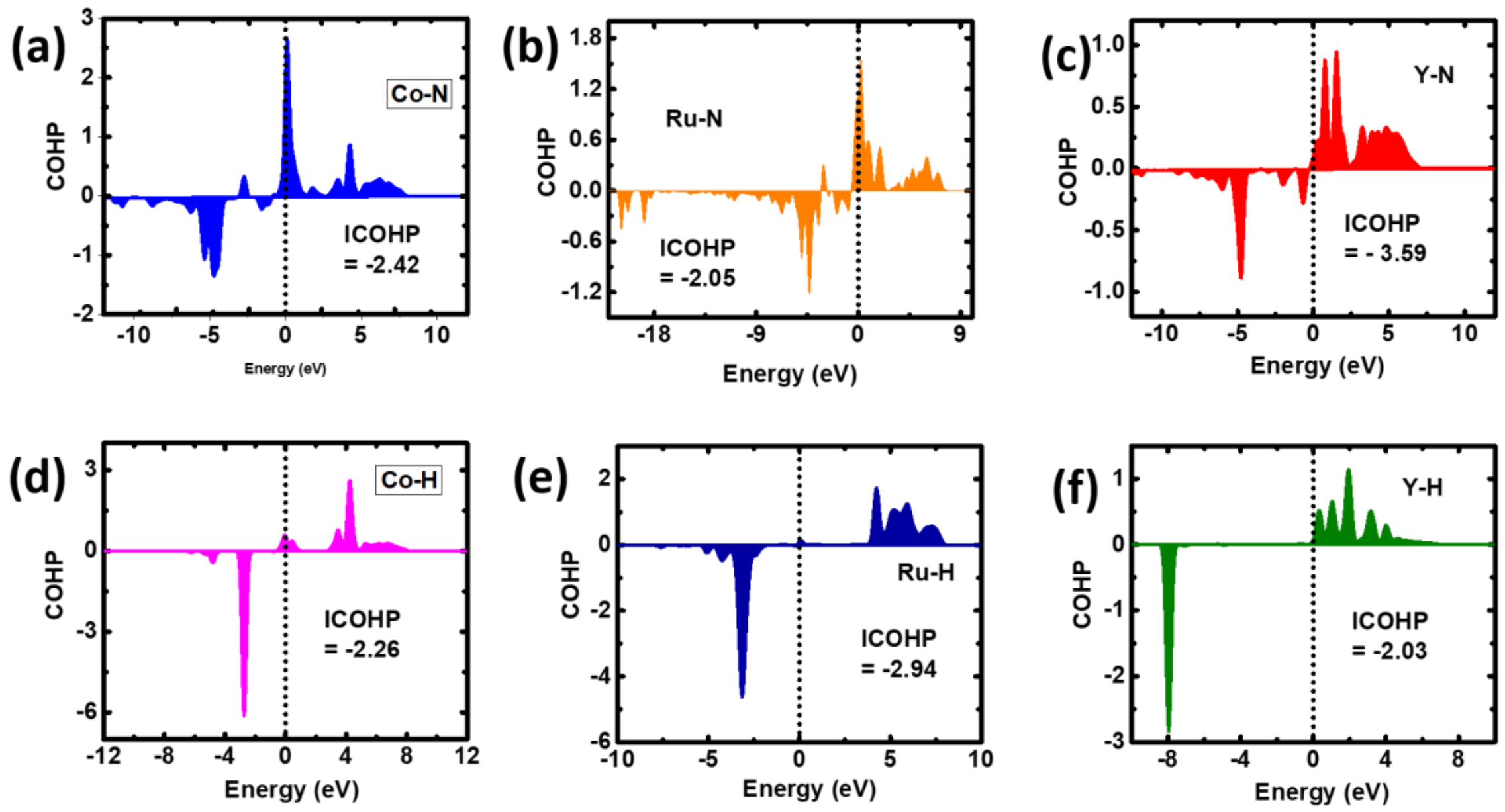

Figure 4: Crystal orbital Hamiltonian population (COHP) analysis for the aza-triphenylene decorated with (a, d) Co, (b, e) Ru and (c, f) Y as a single-atom catalyst.

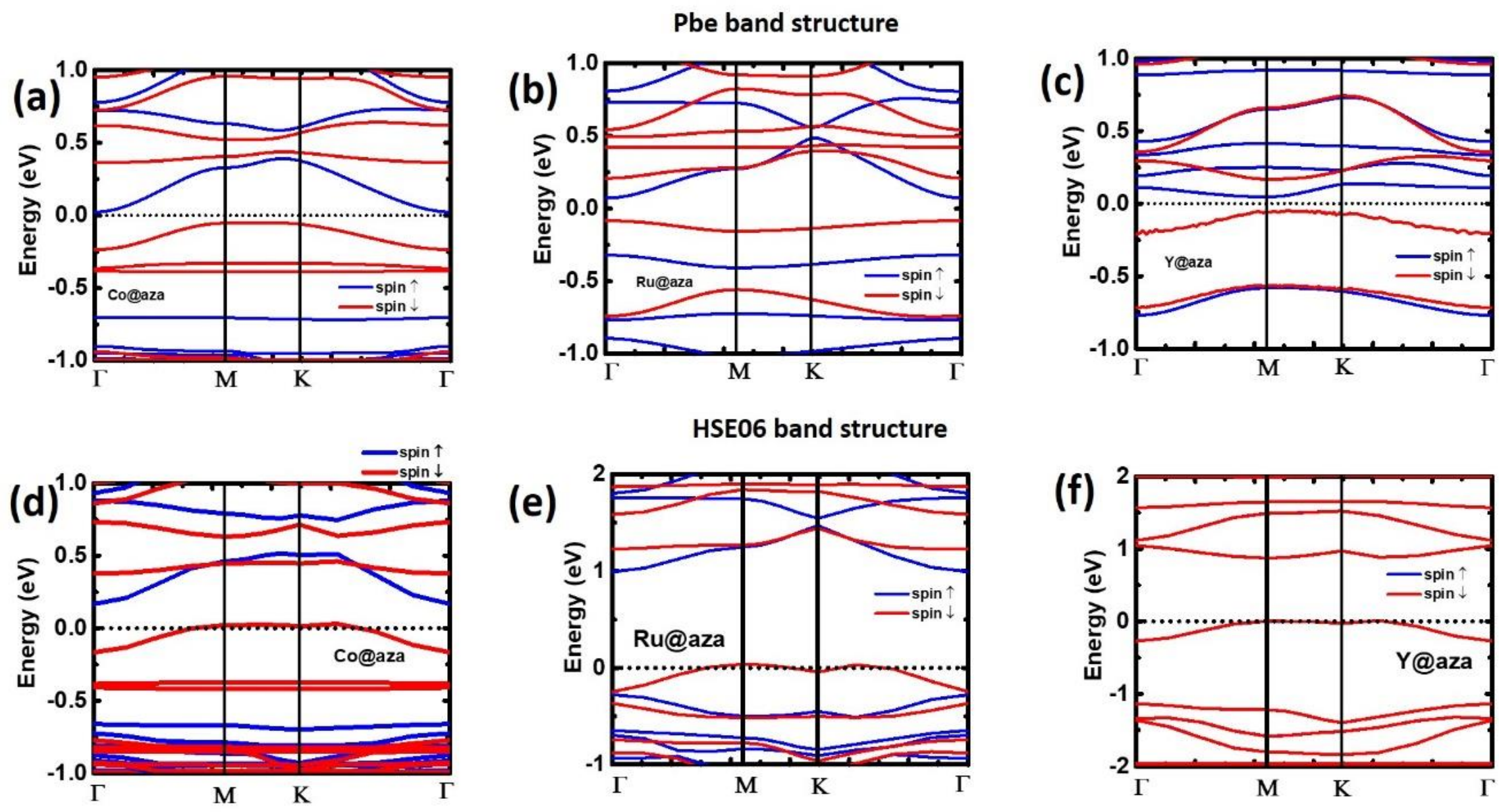

Figure 5: Spin-polarised electronic band structure of aza-triphenylene monolayer adsorbed with (a, d) Co, (b, e) Ru and (c, f) Y as single-atom catalysts. The top and bottom panels indicate the band structure results using PBE-GGA and HSE-06 hybrid functional, respectively. The Fermi level was set to zero, as indicated by the dotted horizontal line.

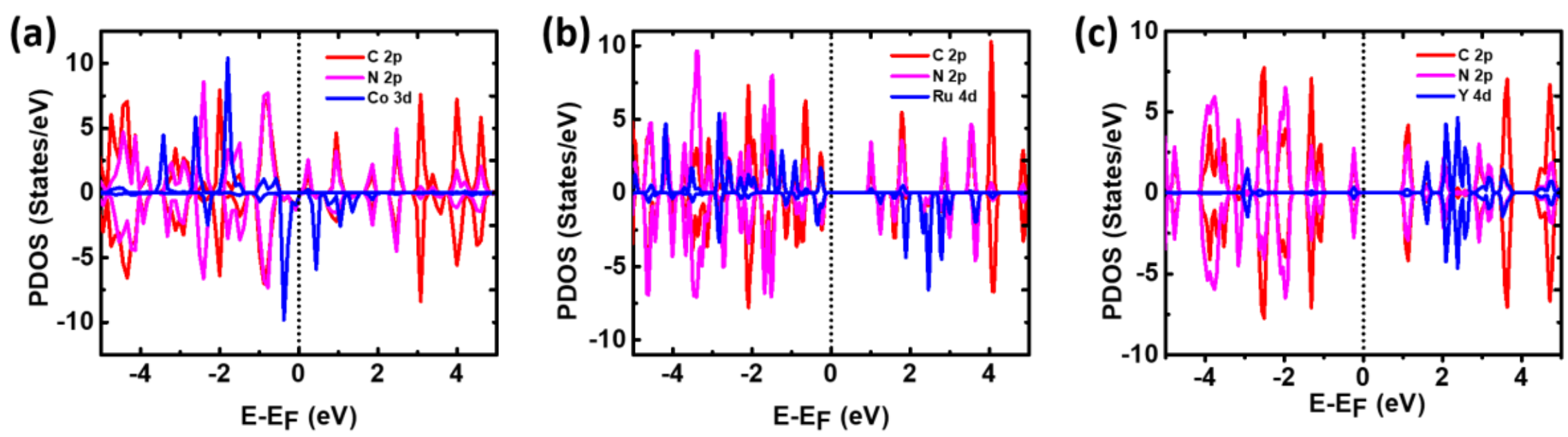

Figure 6: Spin-polarised projected density of states (using HSE-06 functional) of aza-triphenylene monolayer adsorbed with (a) Co, (b) Ru and (c) Y as single-atom catalysts. The Fermi level was set to zero, as indicated by the dotted vertical line.

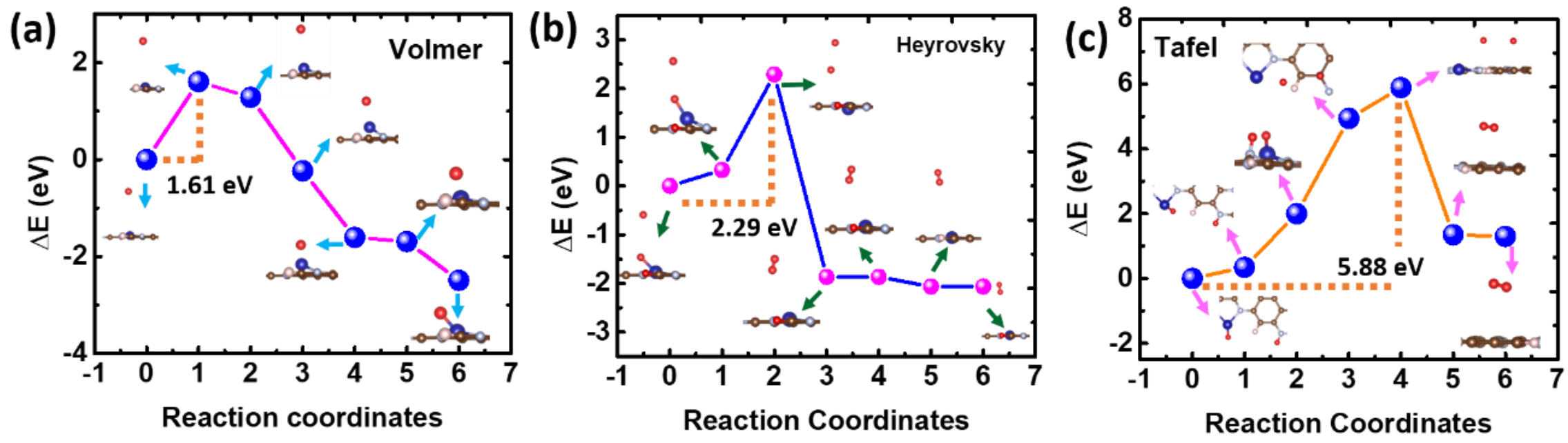

Figure 7: Climbing image nudged elastic band calculation showing the reaction pathway for hydrogen evolution reaction in Co-adsorbed aza-triphenylene (Co@aza) monolayer under (a) Volmer, (b) Heyrovsky and (c) Tafel reaction schemes in the presence of an acidic medium.

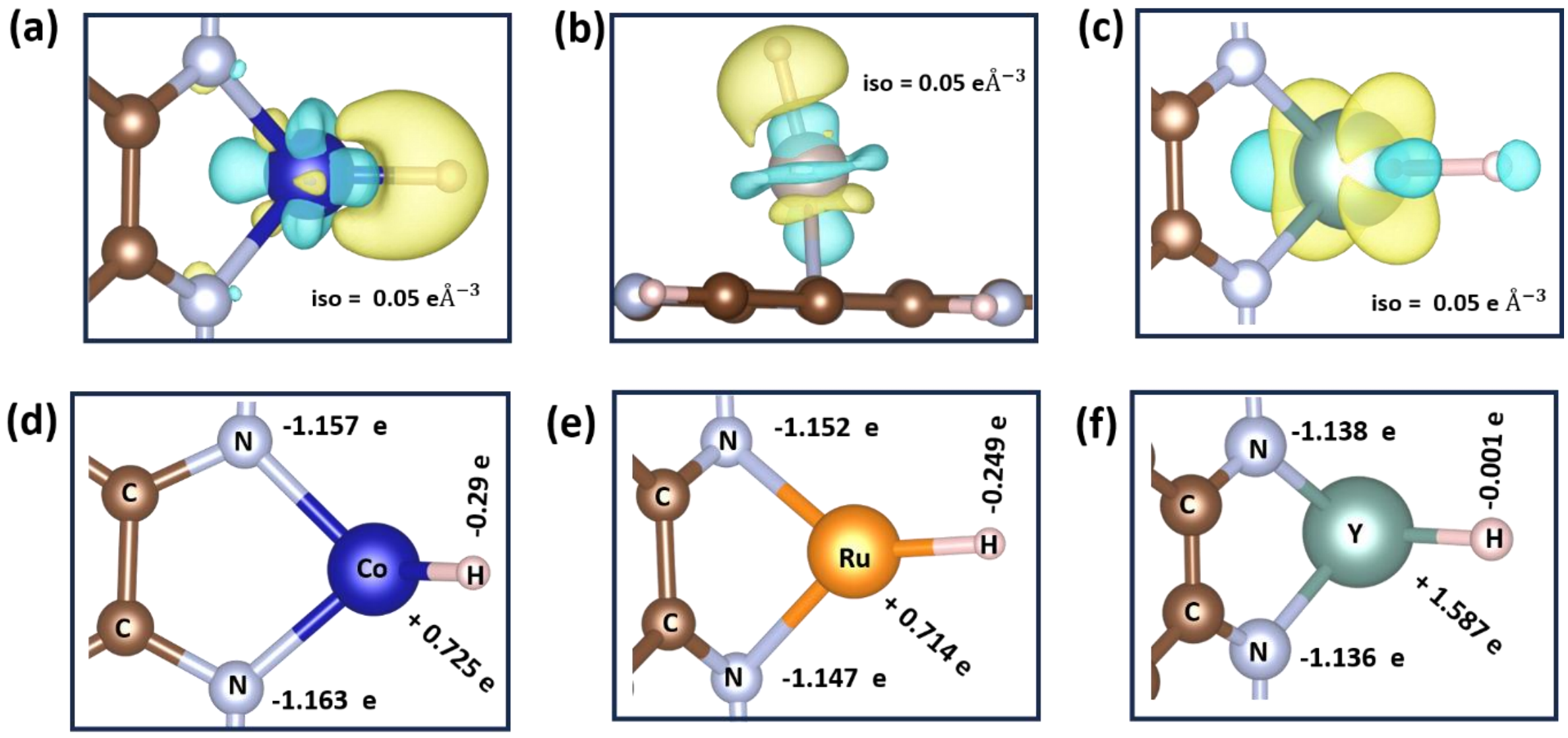

Figure 8: Calculated charge density difference (top panel) and Bader charges (bottom panel) for the aza-triphenylene monolayer adsorbed with Co (a, d), Ru (b, e) and Y (c, f), respectively.

# Supporting Information

## Transition Metal-Anchored Newly Synthesized Aza-Triphenylene Monolayer for Efficient Hydrogen Evolution Reaction: Insights from DFT

Soumendra Kumar Das[1,2], Jagannath Pradhan[2], Prasanjit Samal[3], Sridhar Sahu[2] and Brahmananda Chakraborty[4,5]

[1]*Department of Chemistry and Chemical Engineering, Inha University, Incheon, 22212, Republic of Korea*

[2]*Department of Physics, Indian Institute of Technology (Indian Institute of Mines) Dhanbad, Jharkhand-826004, India.*

[3]*School of Physical Sciences, National Institute of Science Education and Research (NISER) Bhubaneswar, An OCC of Homi Bhabha National Institute, Bhubaneswar, 752050, Odisha, India.*

[4]*High Pressure Crystallography Section, Bhabha Atomic Research Centre (BARC), Trombay, Mumbai 400085, India*

[5]*Homi Bhabha National Institute, Anushaktinagar, Mumbai 400094, India*

Corresponding author: *brahma@barc.gov.in*

| Transition element | Total energy of TM@aza with hydrogen (eV) $E_{aza+TM+H}$ | Total energy of TM@aza without hydrogen (eV) $E_{aza+TM}$ | Total energy of single hydrogen atom (eV) $E_{1/2H}$ | Gibb's free energy charge for HER (eV) $\Delta G_{HER}$ |
|---|---|---|---|---|
| Cr | -399.7747 | -398.6785 | -3.3862 | 2.5389 |
| Mn | -400.6210 | -398.4031 | -3.3862 | 1.4172 |
| Mo | -399.2354 | -398.1199 | -3.3862 | 2.5196 |
| Pd | -396.6149 | -393.3680 | -3.3862 | 0.3882 |
| Pt | -397.0570 | -393.1529 | -3.3862 | -0.2689 |
| Rh | -399.2113 | -394.4212 | -3.3862 | -1.1549 |
| Ru | -398.7616 | -395.3335 | -3.3862 | 0.2071 |
| Ti | -401.2411 | -398.1839 | -3.3862 | 0.5780 |
| V | -399.3648 | -398.9657 | -3.3862 | 3.2361 |
| Y | -401.6798 | -398.1072 | -3.3862 | 0.0625 |
| Zr | -401.8096 | -398.3919 | -3.3862 | 0.2175 |
| Fe | -400.8662 | -396.1418 | -3.3862 | -1.0892 |
| Co | -400.5289 | -396.9518 | -3.3862 | 0.0580 |

Table S1: The calculated Gibbs' free energy change for HER in aza-triphenylene monolayer decorated by a wide range of single transition metals.

| System | Overpotential (eV) | Reference No. |
|---|---|---|
| Ti@$C_2N$ | -0.05 | [1] |
| Mo@$C_2N$ | -0.02 | [1] |
| Ir@Sb | -0.009 | [2] |
| Pt@Sb | -0.010 | [2] |
| Ni@Sb | 0.059 | [2] |
| Cu@Sb | 0.193 | [2] |
| V@$MoS_2$ (Mo site) | -0.16 | [3] |
| Ta@, W@$Fe_2Te_2$ | -0.009, -0.018 | [4] |
| Fe@$Ti_3N_2$ | -0.349 | [5] |
| Mn@, Fe@3N-GY, | 0.081, 0.047 | [6] |

Table S2: Comparison of HER overpotential values for different 2D materials reported in literature.

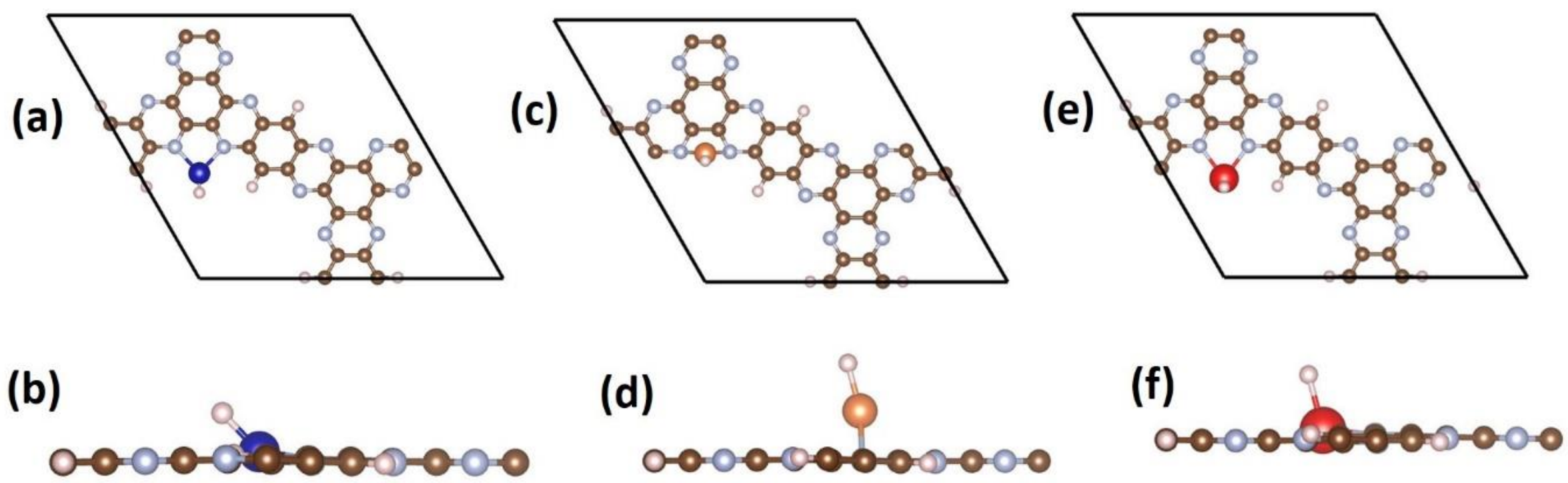

Figure S1: Schematic of the aza-triphenylene monolayer adsorbed with single transition metal such as (a, b) Co, (c, d) Ru, (e, f) Y after geometrical optimization. The top and bottom panel indicates the top view and side views, respectively.

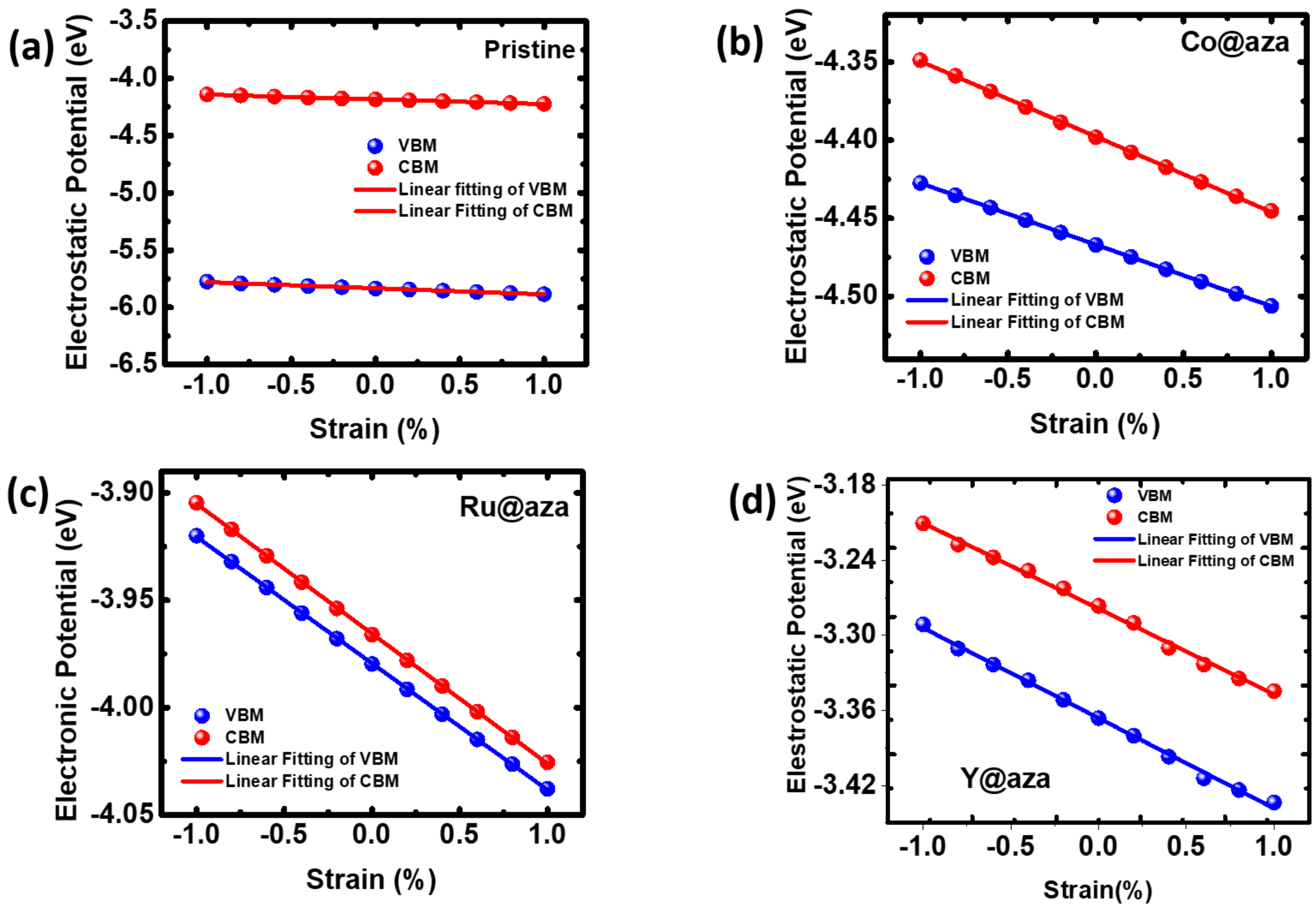

Figure S2: Deformation potential as a function of applied strain for (a) Pristine, (b) Co@, (c) Ru@ and (d) Y@ decorated aza-triphenylene system.

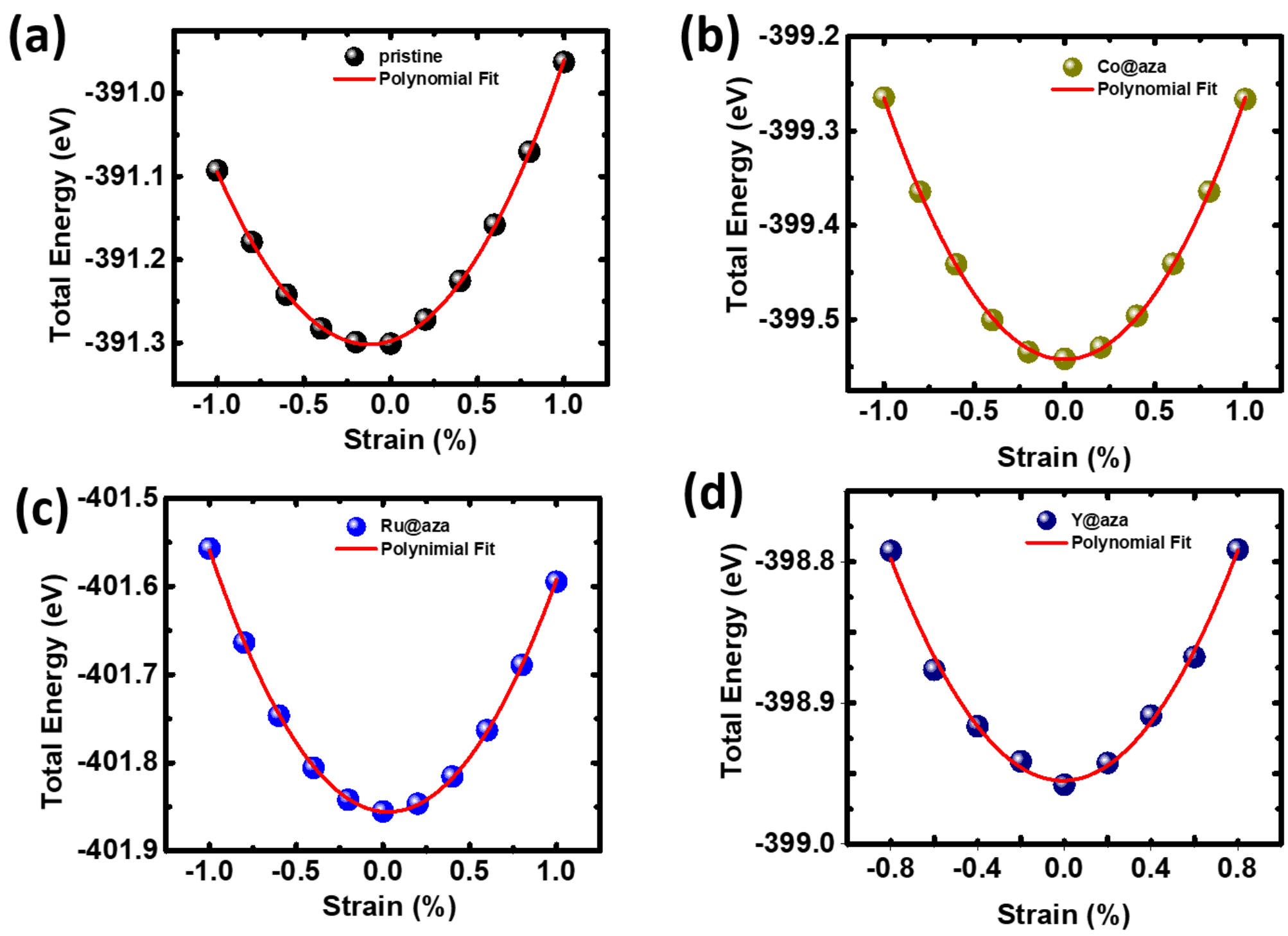

Figure S3: Total energy vs strain curve for (a) Pristine, (b) Co@, (c) Ru@ and (d) Y@ decorated aza-triphenylene system for the calculation of stretching modulus ($C_{2D}$)